\documentstyle[12pt]{report}
\begin{document}
\baselineskip=0.8cm
\ \ \

\begin{center}
{\bf $V$ and $I$ Photometry of Bright Giants in the Central Regions of NGC147}
\end{center}

\begin{center}
T. J. Davidge,

Department of Geophysics and Astronomy,

University of British Columbia, 129-2219 Main Mall,

Vancouver, BC CANADA V6T 1Z4

and

Gemini Canada Project Office, Dominion Astrophysical Observatory,

5071 W. Saanich Road, Victoria, BC, CANADA V8X 4M6$^a$

e-mail: davidge@dao.nrc.ca
\end{center}

\vspace{1.0cm}
\begin{center}
To appear in The Astronomical Journal
\end{center}

\vspace{1.0cm}
\noindent{$^a$ Postal Address

\pagebreak[4]
\begin{center}
{\bf ABSTRACT}
\end{center}

	Deep $V$ and $I$ CCD images with sub-arcsec spatial resolution are used
to investigate the stellar content of the central regions of the Local Group
dwarf elliptical galaxy NGC147. Red giant branch (RGB) stars are resolved over
the entire field, and the RGB-tip occurs at $I \sim
20.5$, suggesting that the distance modulus is 24.3. A
comparison with globular cluster sequences indicates that the center
of NGC147 is moderately metal-poor, with $\overline{[Fe/H]} \sim -1$. This
is not significantly different from what was found in the outer regions of
the galaxy by Mould, Kristian \& Da Costa (1983, ApJ, 270, 471).
Moreover, the width of the $V-I$ color distribution at $I = 21.0$
indicates that a spread in metallicity is
present, with $\sigma_{[Fe/H]} \sim \pm 0.3$. There is no evidence of a
component more metal-poor than [Fe/H] $\sim -1.3$. A small population of
moderately bright asymptotic giant branch (AGB) stars has also been
detected, and the AGB-tip occurs near M$_{bol} \sim -5.0$, indicating
that an intermediate-age population is present. It is estimated that the
intermediate-age population contributes $\sim 2 - 3\%$ of the $V$ light from
NGC147.

\pagebreak[4]
\begin{center}
1. INTRODUCTION
\end{center}

	Because they are relatively close, Local Group galaxies provide unique
templates for interpreting more distant, unresolved, objects. Studies of
nearby dwarf galaxies are of particular interest, as
it has recently been suggested that low mass systems
at intermediate redshifts may have experienced large bursts of
star formation (e.g. Broadhurst, Ellis, \& Shanks 1988; Eales 1993; Colless et
al. 1993). One signature of large-scale star formation at moderate
redshifts will be a population of intermediate-age stars, and the spatial
distribution of such a component may influence
present-day morphologies. Indeed, Babul \& Reese (1992) suggest that
non-nucleated dwarf elliptical (dE) and nucleated dwarf elliptical (dE,n)
systems may have experienced systematically different star-forming histories.
Comparative studies of the stellar contents of nearby nucleated and
non-nucleated dwarf ellipticals provide a direct means of testing predictions
of this nature.

	A potential problem with using Local Group galaxies as models
for more distant systems is that many are companions to the Galaxy or M31,
so that tidal interactions may have influenced their evolution.
However, early-type dwarf galaxies tend to cluster
around more massive systems (e.g. Vader \& Sandage 1991), suggesting that the
companions to M31 and the Galaxy may actually provide more `typical' templates
for dwarf galaxy evolution than isolated members of the Local Group. The M31
satellite system provides a number of lucrative targets for detailed
investigation, which cover a range of
morphological types. Three of the satellites of M31 are low surface-brightness
dwarf ellipticals: NGC147, NGC185, and NGC205. NGC205 is type dE,n while
NGC147 and NGC185 are type dE (e.g. Kent 1987).
These galaxies have conspicuously different young stellar
contents. Early photographic studies of NGC185 and NGC205 by Baade (1951) and
Hodge (1963; 1973) revealed centrally-concentrated populations of blue
stars. Moreover, NGC185 and NGC205 both contain dust clouds (Hodge
1963, 1973), suggesting that star-forming activity may not have ceased.
On the other hand, NGC147 does not contain massive young stars, and there is
no evidence for dust absorption (Hodge 1976), suggesting that massive stars
will not form in the immediate future.

	Although relatively easy to detect, massive stars probe only the most
recent histories of galaxies. Observations which sample longer time spans,
preferably extending into intermediate epochs, will provide a more thorough
means of comparing the evolutionary histories of dwarf
galaxies. Deep red and near-infrared images are particularly
effective for detecting bright AGB stars, and observations of this
nature have revealed intermediate-age populations in NGC185 and NGC205 (Richer,
Crabtree \& Pritchet 1984, Davidge 1992, Lee, Freedman, \& Madore 1993). At
present, there is no evidence to suggest that an intermediate age population is
present in NGC147. Mould, Kristian, \& Da Costa (1983) obtained deep $V$ and
$I$ images of a field 5 arcmin from the galaxy center and found a
well-defined red giant branch (RGB) with a color appropriate
for [Fe/H] $\sim -1.2$. Bright AGB stars were abscent, and a
subsequent study of star clusters by Da Costa \& Mould (1988) also failed to
detect an intermediate-age population.

	If present in NGC147, luminous AGB stars should be
most easily detected in the central regions of the galaxy, where the stellar
density is greatest, and the chances of detecting stars during short-lived
phases of their evolution is largest. In an effort to
search for luminous AGB stars, a series of deep
$V$ and $I$ images were obtained of the center of NGC147,
and the results of this survey are reported here. The data
and their reduction are discussed in Section 2. The $(I, V-I)$ color-magnitude
diagram (CMD) and $I$ luminosity function are presented
in Section 3, and these reveal a well-defined RGB, with a modest population
of AGB stars extending up to M$_{bol} \sim -5.0$. In Section 4
the $V-I$ color distribution at $I \sim 21.0 \pm 0.25$ is examined in an effort
to determine the mean metallicity and metallicity dispersion. A brief summary
and discussion of the results follows in Section 5.

\begin{center}
2. OBSERVATIONS AND REDUCTIONS
\end{center}

	The data were recorded during the night of UT January 1/2 1992 using
the HRCAM imager (McClure et al. 1989), mounted at the prime focus of the
3.6 metre Canada-France-Hawaii Telescope (CFHT). The detector was SAIC1, a
thick Ford-Aerospace type CCD with 18$\mu$m
square pixels in a 1024 x 1024 format. The spatial scale produced by
this instrument/detector combination is 0.13 arcsec/pixel, so that each
image covered $\sim 5$ square arcmin.

	A field centered near right ascension $00^{h}
30^{m} 32^{s}$ and declination $+48^{o} 13^{'}
37^{"}$ (epoch 1950) was observed. Six 300 sec exposures were recorded
through a Kron-Cousins $I$ filter, while ten 300 sec exposures were recorded
in $V$. The telescope pointing experienced a $\sim 25$ arcsec jump mid-way
through the observing sequence, which went undetected at the telescope. Hence,
the field coverage of the final combined images is smaller than would otherwise
have been the case ($\sim 2.5$ square arcmin). The mean seeing, derived from
bright stars in the final combined images, was 0.8 arcsec FWHM in both filters.

	The data reduction followed standard lines. A median bias frame
was subtracted from the raw exposures, and the results were divided by dome
flats appropriate for each filter. Corrections were not made for detector dark
current due to the relatively short exposure times. The flat-fielded images
were aligned, co-added, and trimmed to the area common to all exposures. The
final $I$ image is shown in Figure 1.

\begin{center}
3. PHOTOMETRY
\end{center}

\noindent{\it 3.1 Stellar Brightnesses}

	Stellar brightnesses were measured with the PSF-fitting
routine ALLSTAR (Stetson \& Harris 1988), which is part of the DAOPHOT
(Stetson 1987) photometry package. Standard stars in the globular
cluster NGC2419 (Christian et al. 1985) were used to calibrate
the photometry. Standards were observed only once during the night,
so mean extinction coefficients, derived from previous
runs, were adopted for calibration purposes. The uncertainties in the
photometric zeropoints are $\pm 0.02$ magnitudes.

	Artificial star experiments were run to determine completeness
fractions and assess systematic effects in the photometry.
The results depend on crowding, the extent of which
changes across the field, being greatest near the center of the galaxy. To
partially compensate for spatial variations in stellar density, the field was
divided into three regions, centered on the globular cluster Hodge 1, which
lies close to the photometric center of NGC147 (Hodge 1976). The boundaries of
these regions were selected to produce equal integrated brightnesses; hence,
to first order the regions should contain comparable numbers of stars. Region 1
covers a 27 arcsec radius centered on Hodge 1, while Region 2 spans the
interval from 27 to 43 arcsec. Region 3 covers the remaining parts of
the image. The total areas covered on the final images are 0.45, 0.55, and
1.53 square arcmin, respectively. The data are 50\% complete in Regions 1 , 2,
and 3 when $V = 22.6, 22.9,$ and 23.1, and $I = 22.0, 22.2,$ and 22.5.

	Hodge (1976) measured the brightness of the cluster Hodge 1
photoelectrically, and found that $V = 17.66$ within a 14 arcsec aperture.
For comparison, the current data indicate that $V = 17.57$ and $V-I = 1.50$
within the same aperture. These values may not be representative of the
cluster, as the CCD data reveal that the light profile of
Hodge 1 declines at a relatively steep rate, such that at a 2 arcsec radius the
mean $I$ surface brightness is comparable to that of the surrounding galaxy.
Indeed, $V-I$ becomes steadily bluer as aperture size decreases,
such that with a 2 arcsec aperture $(V-I) \sim 1.15$, while $V \sim 18.68$.
Adopting $E(B-V) \sim 0.18$ (Burstein \& Heiles 1984),
a value which will be used throughout this paper, then the reddening
curve of Dean, Warren, \& Cousins (1978)
implies that $E(V-I) = 0.23$, so that $(V-I)_0 = 0.92$ within the 2 arcsec
aperture. Da Costa \& Mould (1988) studied the spectroscopic properties
of Hodge 1, and found that the W(K) $+$
W(M) index, where W(K) and W(M) measure the strengths of Ca K and red metallic
features, respectively, is similar to M13, for which $(V-I)_0 \sim 0.8$ (Reed,
Hesser, \& Shawl 1988). Hence, there is broad agreement between
the spectroscopic and photometric properties of Hodge 1.

\vspace{0.5cm}
\noindent{\it 3.2 The CMD's and Luminosity Functions}

	The $(I, V-I)$ diagram of the entire field is shown in Figure 2, and
the red giant branch (RGB) is clearly visible. A population of moderately
bright red stars, which are evolving on the AGB, is also apparent, and
the upper envelope of this sequence occurs near $I \sim 19.8$. The
small number of stars brighter than $I \sim 19.0$ in Figure 2 are
probably foreground objects. Evidence for this comes from the Galactic
star count models of Ratnatunga \& Bahcall (1985), who predict that
$\sim 4$ stars brighter than $V = 19$ should be detected in the current field.
For comparison, 6 stars brighter than $V = 19$ were detected, in
reasonable agreement with the model star counts.

	The $(I, V-I)$ CMD's for Regions 1, 2, and 3 are compared in Figure 3.
Region 1 is more crowded than Region 3, so it is not surprising that more
stars were detected in the latter area than the former. If the central regions
of NGC147 are dominated by an old population, as appears to be the case
(Section 5), then the mean metallicity can be determined by comparison
with Galactic globular cluster sequences. The loci of the clusters NGC7078
and 47 Tuc ([Fe/H] $\sim -2.2$ and $-0.7$; Zinn \& West 1984), as tabulated
by Da Costa \& Armandroff (1990), are compared
with the NGC147 observations in Figure 3. A distance modulus of 24.3,
which places the RGB-tip at $I \sim 20.5$ (see below) if $E(V-I) \sim 0.23$,
has been assumed. It is evident that the majority of stars in all three
regions fall between the NGC7078 and 47 Tuc loci, although there is a small
spray of stars which fall redward of the 47 Tuc sequence. The nature of these
objects will be discussed further in Section 4.

	The $I$ luminosity function can be used to determine the
brightness of the RGB-tip. A large dataset is required for this task, as the
RGB-tip may be a relatively small feature, superimposed on a sloping
luminosity function. In order to use the maximum possible
number of stars from the current dataset, a composite $I$ luminosity function
for all three regions was constructed using all stars detected in $I$,
and the result is shown in Figure 4. In this Figure $N$ is
the number of stars per 0.25 magnitude per square arcmin. A discontinuity
is apparent near $I \sim 20.5$, which is due to the RGB-tip. For comparison,
Mould et al. (1983) concluded that the RGB-tip occurs at
$I \sim 20.4$ in NGC147.

	The absence of luminous resolved early-type stars suggests that, unlike
NGC205 and NGC185, there has not been recent star formation in NGC147.
However, the brightness of the AGB-tip suggests that star formation did
occur during intermediate epochs. The brightness of the AGB-tip
is best determined from a bolometric, rather than monochromatic, luminosity
function, as the AGB sequence becomes degenerate with color at high
luminosities. The brightest AGB stars in an old, moderately metal-poor stellar
system, such as the globular cluster 47 Tuc, have M$_{bol} \sim -4.5$ (Frogel,
Persson, \& Cohen 1981). Consequently, if the AGB-tip in NGC147 is brighter
than this then it is likely that an intermediate-age component is present.
A bolometric AGB luminosity function for NGC147 has been computed using the
procedure described by Reid \& Mould (1984), with bolometric corrections
derived from Equation 1 of Bessell \& Wood (1984). The result,
corrected for completeness, is shown in Figure 5, where $N$ is the number
of stars per square arcmin per 0.25 magnitude interval. A distance modulus
$\mu \sim 24.3$ has been assumed. The brightest stars have M$_{bol} \sim
-5.0$, indicating that an intermediate age component is present.

\begin{center}
4. THE RGB COLOR DISTRIBUTION
\end{center}

	The color distribution of RGB stars can be used to determine
the mean metallicity and metallicity dispersion
near the center of NGC147. The distribution of $V-I$ colors for stars
with $I \sim 21.0 \pm 0.25$ in Regions 1, 2, and 3 are compared in Figure 6.
These distributions were computed using a 0.2 magnitude bin in $V-I$, and
have been corrected for incompleteness.
The mean colors in the three zones are very similar, with $\overline{V-I} \sim
2.0$. Moreover, in all three cases there is a steep blue
envelope at $V-I \sim 1.4$, and a red tail
which extends beyond $V-I \sim 2.0$. The Region 1 and 2 distributions show a
two-peaked profile not apparent in Region 3; however, a Kolmogoroff-Smirnoff
test indicates that the three distributions are not significantly different.

	The composite color distribution for all three fields
is shown in Figure 7. A series of model color distributions,
broadened with error functions determined from the artificial star experiments,
were computed using the globular cluster loci tabulated by Da Costa \&
Armandroff (1990), and the results for
NGC1851 ([Fe/H] $\sim -1.3$; Zinn \& West 1984) and 47 Tuc are compared with
the observations in Figure 7. No attempt was made to produce optimum fits
with the observations; rather, the models were simply scaled so that
their peaks matched the observed color distributions. It is apparent that (1)
a range of metallicities is required to match the color distribution;
(2) the majority of stars have [Fe/H] between that of NGC1851 and 47 Tuc,
so that $\sigma_{[Fe/H]} \sim \pm 0.3$, and $\overline{[Fe/H]} \sim -1$; (3)
there is no evidence for a very metal-poor (ie. [Fe/H] $\sim -2$) population;
and (4) the spray of stars falling redward of the 47 Tuc sequence in Figure
3 is actually the gaussian tail of a moderately metal-poor component.
The mean metallicity found here is consistent with that derived by Mould et
al. (1983) at larger radii. Mould et al. (1983) also found evidence for a
metallicity dispersion, as did Saha, Hoessel, \& Mossman (1990), who studied
the period distribution of RR Lyrae variables.

\begin{center}
5. SUMMARY AND DISCUSSION
\end{center}

	The photometry presented in this paper indicates that the central
regions of NGC147 are moderately metal-poor, and
contain stars spanning metallicities in the range $-1.3 \leq
[Fe/H] \leq -0.7$. Moreover, a small population of stars which formed at
intermediate epochs is also present. The implications of these results are
briefly discussed below.

	The $I, V-I$ CMD in Figure 2 is morphologically similar to that
constructed by Mould et al. (1983) for a field 5 arcmin from the
center of NGC147, and the current results indicate that mean metallicity
does not vary significantly across the face of the galaxy.
Consequently, if a metallicity gradient is present in NGC147 then it must be
very shallow. This is qualitatively consistent with non-conservative
galaxy formation models, which predict that supernovae-driven winds eject
star-forming material from low mass systems early in their evolution,
halting chemical enrichment and preventing dissipative collapse (e.g.
Carlberg 1984; Dekel \& Silk 1986). The seeming absence of a very low
metallicity (ie. [Fe/H] $\sim -2$) component suggests that the main body
of NGC147 may have experienced very rapid initial chemical enrichment or formed
from previously enriched material. However, NGC147 is not completely
devoid of metal-poor stars, as Da Costa \& Mould (1988) found that the
clusters Hodge 1 and Hodge 3 have [Fe/H] $\sim -2.0$.

	The age of the intermediate age component in NGC147 can be estimated
from the brightness of the AGB-tip. Using the calibration of Mould \& Aaronson
(1988), M$_{bol} \sim -5$ corresponds to an age of $\sim 5$ Gyr. The detection
of intermediate-age stars in the central regions of NGC147 is of interest as
Mould et al. (1983) failed to detect bright AGB stars 5 arcmin from the
galaxy center. However, this may be a consequence of the low density of these
objects at large radii. This can be demonstrated by comparing the mean surface
brightnesses in the central and Mould et al. (1983) fields. The luminosity
profile measured by Kent (1987) indicates that the mean surface brightness in
the Mould et al. field is $\mu_g \sim 26$ mag/arcsec$^2$, compared with $\mu_g
\sim 21.7$ mag/arcsec$^2$ near the galaxy center. Therefore,
the stellar density in the Mould et al. field should be $\sim 0.02$ times that
near the center. After correcting for differences in field size,
Mould et al. (1983) should have detected one tenth
the number of bright AGB stars found in the current study
{\it if the relative densities of old and intermediate-age stars were the same
as in the galaxy center}. Roughly 50 stars brighter than the RGB-tip have been
detected in the current study, so $\sim 5$ would be expected in the
Mould et al. (1983) field. Such a small number of stars may be difficult to
detect given the foreground contamination towards NGC147 (eg. Figure 9 of
Mould et al. 1983).

	The contribution that the intermediate-age component
makes to the integrated light from NGC147 can be estimated from
the AGB luminosity function. Frogel, Mould, \& Blanco (1990) computed
the fractional contribution that AGB stars make to
total cluster luminosities. Unfortunately, this calibration uses AGB
stars as faint as M$_{bol} \sim -3.6$, and so is unsuitable for the
current study, as NGC147 is likely not coeval, and AGB stars fainter than
M$_{bol} \sim -4.5$ could belong to older populations. Consequently, a revised
calibration was derived from the Frogel et al. (1990) dataset using only AGB
stars with M$_{bol} \leq -4.5$. Moreover, the sample was restricted to
clusters of SWB (Searle, Wilkinson, \& Bagnuolo 1980) types 5.5 and 6,
as these have ages comparable to the intermediate-age population in
NGC147 (e.g. Table 3 of Frogel et al. 1990). This exercise revealed that
AGB stars brighter than M$_{bol} \sim -4.5$ contribute $0.22 \pm 0.05$ of the
total luminosity from SWB types 5.5 and 6 clusters. A direct integration
of the AGB luminosity functions from Regions 1, 2, and 3
reveals that the total luminosity of bright AGB stars is M$_{bol} \sim -6.6$,
so that the total luminosity of the intermediate-age population near the center
of NGC147 is M$_{bol} \sim -8.2$. If the spectral-energy distribution of this
population is similar to that of a late G or early K giant then the bolometric
correction will be $\sim - 0.1$, so that M$_V \sim -8.3$. The integrated
brightness of the current field is M$_V \sim -12.3$. Hence, the
intermediate-age component contributes only $\sim 2 - 3\%$ of the total light
output in $V$ near the galaxy center.

\vspace{0.5cm}
	It is a pleasure to thank the Natural Sciences and Engineering
Research Council of Canada (NSERC) and the National Research Council of
Canada (NRC) for financial support. Sincere thanks are also extended to
Sidney van den Bergh for commenting on an earlier version of this paper.
An anonymous referee also made comments which revealed a calibration error,
and greatly improved the presentation of the paper.

\pagebreak[4]
\begin{center}
REFERENCES
\end{center}

\parindent=0.0cm

Baade, W. 1951, Pub. Univ. Michigan Obs., 10, 7

Babul, A., \& Rees, M. J. 1992, MNRAS, 255, 346

Bessell, M. S., \& Wood, P. R. 1984, PASP, 96, 247

Broadhurst, T. J., Ellis, R. S., \& Shanks, T. 1988, MNRAS, 235, 827

Burstein, D., \& Heiles, C. 1984, ApJS, 54, 33

Carlberg, R. G. 1984, ApJ, 286, 403

Christian, C. A., Adams, M., Barnes, J. V., Butcher, H., Hayes, D. S.,
\linebreak[4]\hspace*{2.0cm}Mould, J. R., \& Siegel, M. 1985, PASP, 97, 363

Colless, M., Ellis, R. S., Broadhurst, T. J., Taylor, K., \& Peterson,
\linebreak[4]\hspace*{2.0cm}B. A. 1993, MNRAS, 261, 19

Da Costa, G. S., \& Armandroff, T. E. 1990, AJ, 100, 162

Da Costa, G. S., \& Mould, J. R. 1988, ApJ, 334, 159

Davidge, T. J. 1992, ApJ, 397, 457

Dean, J. F., Warren, P. R., \& Cousins, A. W. J. 1978, MNRAS, 183, 569

Dekel, A., \& Silk, J. 1986, ApJ, 303, 39

Eales, S. 1993, ApJ, 404, 51

Frogel, J. A., Mould, J., \& Blanco, V. M. 1990, ApJ, 352, 96

Frogel, J. A., Persson, S. E., \& Cohen, J. G. 1981, ApJ, 246, 842

Hodge, P. W. 1963, AJ, 68, 691

Hodge, P. W. 1973, ApJ, 182, 671

Hodge, P. W. 1976, AJ, 81, 25

Kent, S. M. 1987, AJ, 94, 306

Lee, M. G., Freedman, W. L., \& Madore, B. F. 1993, AJ, 106, 964

McClure, R. D., Grundmann, W. A., Rambold, W. N., Fletcher, J. M.,
\linebreak[4]\hspace*{2.0cm} Richardson,
E. H., Stilburn, J. R., Racine, R., Christian,
\linebreak[4]\hspace*{2.0cm}C. A., \& Waddell, P. 1989, PASP, 101, 1156

Mould, J. R., \& Da Costa, G. S. 1988, in Progress and Opportunities in
\linebreak[4]\hspace*{2.0cm}Southern Hemisphere Optical Astronomy, ed. V. M.
Blanco \& M.\linebreak[4]\hspace*{2.0cm} M. Phillips (ASP: San Francisco), 197

Mould, J. R., Kristian, J., \& Da Costa, G. S. 1983, ApJ, 270, 471

Ratnatunga, K. U., \& Bahcall, J. N. 1985, ApJS, 59, 63

Reed, E. C., Hesser, J. E., \& Shawl, S. J. 1988, PASP, 100, 545

Reid, N., \& Mould, J. 1984, ApJ, 284, 98

Richer, H. B., Crabtree, D. R., \& Pritchet, C. J. 1984, ApJ, 287, 138

Saha, A., Hoessel, J. G., \& Mossman, A. E. 1990, AJ, 100, 108

Searle, L., Wilkinson, A., \& Bagnuolo, W. G. 1980, ApJ, 239, 803

Stetson, P. B. 1987, PASP, 99, 191

Stetson, P. B., \& Harris, W. E. 1988, AJ, 96, 909

Vader, J. P., \& Sandage, A. 1991, ApJL, 379, L1

Zinn, R., \& West, M. J. 1984, ApJS, 55, 45

Zinnecker, H., \& Cannon, R. D. 1985, in Star Forming Dawrf Galaxies and
\linebreak[4]\hspace*{2.0cm}Related Objects, ed. D. Kunth, T. X. Thuan, \& J.
Tran\linebreak[4]\hspace*{2.0cm}Tranh Van (Gif-Sur-Yvettes:Editions
Frontieres), 155

\pagebreak[4]
\begin{center}
FIGURE CAPTIONS
\end{center}

FIG. 1. The final $I$ image of the NGC147 field.

\vspace{0.3cm}
FIG. 2. The $(I, V-I)$ CMD of the entire field.

\vspace{0.3cm}
FIG. 3. The $(I, V-I)$ CMD's for, from left to right, Regions 1, 2, and 3. The
solid lines are the loci of the clusters NGC7078 and 47 Tuc from Da Costa \&
Armandroff (1990). The placement of the cluster loci assume
$\mu \sim 24.3$ and $E(B-V) \sim 0.17$.

\vspace{0.3cm}
FIG. 4. The completeness-corrected $I$ luminosity function
for all three regions derived from all stars detected in $I$. Note the
discontinuity near $I \sim 20.5$. $N$ is the number of stars
per 0.25 magnitude interval per square arcmin. The error bars reflect
counting statistics.

\vspace{0.3cm}
FIG. 5. The completeness-corrected bolometric AGB luminosity function for
NGC147. $N$ is the number of stars per 0.25 magnitude interval per square
arcmin. The error bars reflect counting statistics.

\vspace{0.3cm}
FIG. 6. The $(V-I)$ color distributions at $I = 21 \pm 0.25$ for, from top to
bottom, Regions 1, 2, and 3. $n$ is the number of stars per 0.2 magnitude
interval normalised to the total number of stars detected in each region. The
curves have been corrected for incompleteness.

\vspace{0.3cm}
FIG. 7. The composite $V-I$ color distribution at $I \sim 21.0 \pm 0.25$ for
all three regions. Also shown are the color distributions expected for
single-metallicity populations corresponding to the globular clusters
NGC1851 (dashed line) and 47 Tuc (dotted-dashed line). The model
distributions have been scaled to roughly match the observed color
distribution near the peak model distribution.
\end{document}